\documentclass [12pt,a4paper]{article}

\title{Black holes -- Estimation of their lower and upper mass limits
  stemming from the model of Expansive Nondecelerative Universe}
\author{Jozef \v{S}ima and Miroslav S\'{u}ken\'{\i}k \\[1ex]
  Slovak Technical University, FCHPT, Radlinsk\'{e}ho 9, \\
  812 37 Bratislava, Slovakia
  \\
  e-mail: sima@chtf.stuba.sk, sukenik@minv.sk} \date{}
\begin{document}
\maketitle
\begin{abstract}
In this contribution, the thermodynamics of black holes is treated by the 
model of Expansive Nondecelerative Universe (ENU). Based on entropy 
considerations and localization of gravitational energy, estimation of both 
the lower and upper mass limits of black holes is given. Both mass limits 
are time dependent. 


\end{abstract}

\section{Thermodynamics of black holes and a lower limit of their mass}
As part of his famous ideas concerning the existence and properties of black 
holes, combining quantum mechanics and thermodynamics, Hawking [1] 
formulated the following equation describing the quantum evaporation output 
$P_{evap} $ of a black hole with the gravitational radius $r_{BH} $
\begin{equation}
\label{eq1}
P_{evap} = {\frac{{\hbar .c^{2}}}{{r_{BH}^{2}}} }
\end{equation}
Hypothesis on the quantum evaporation of black holes has been developed by 
several groups of theoreticians and field researchers, no burst proving a 
fatal cessation of either primordial or later formed black hole has been, 
however, observed. The questions of minimum and maximum mass of black holes 
are still under discussion and deserve attention. We are convinced that the 
model of Expansive Nondecelerative Universe (ENU) [2-4] is capable to 
contribute in offering answers to such questions. In this contribution its 
potential in this field is documented.

Due to the Vaidya metric application [5], ENU enables to localize and 
quantify the energy of the gravitational field. For weak field conditions it 
adopts the form
\begin{equation}
\label{eq2}
\varepsilon _{g} = - {\frac{{R.c^{4}}}{{8\pi .G}}} = - 
{\frac{{3m.c^{2}}}{{4\pi .a.r^{2}}}}
\end{equation}
where $\varepsilon_g$ is the density of gravitational field energy 
generated by a body with a mass $m$ at a distance $r$, $R$ is the scalar 
curvature and $a$ is the gauge factor (reaching at present $a \cong 
1.3\times 10^{26}$m) [2].

It is worth mentioning that the magnitude of $\varepsilon _{g} $ given
by~(\ref{eq2}) is closed to the $\theta _{0}^{0} $ component of the
Einstein energy-momentum pseudotensor describing the density of
gravitational energy also for strong field conditions [6]. As a direct
consequence of Vaidya metric application, $R \ne 0$ also outside a
body. In such a case the gravitational output $P_{g} $ of a body at a
given cosmological time $t_{c} $ is as follows
\begin{equation}
\label{eq3}
P_{g} = - {\frac{{d}}{{dt}}}\int {{\frac{{R.c^{4}}}{{8\pi .G}}}dV = - 
{\frac{{m.c^{2}}}{{t_{c}}} } = - {\frac{{m.c^{3}}}{{a}}}} 
\end{equation}
In order not to violate the second law of thermodynamics, it must hold
\begin{equation}
\label{eq4}
{\left| {P_{g}}  \right|} \ge P_{evap} 
\end{equation}
In a limiting case of the outputs being equal and taking into account that 
\begin{equation}
\label{eq5}
r_{BH} = {\frac{{2G.m_{BH}}} {{c^{2}}}}
\end{equation}
a lower mass limit of a black hole can be expressed as 
\begin{equation}
\label{eq6}
m_{BH(\min )} = \left( {{\frac{{\hbar .c^{3}.a}}{{4G^{2}}}}} \right)^{1 / 
3}
\end{equation}
that leads to its current value
\begin{equation}
\label{eq:7a}
m_{BH(\min )} \cong 10^{12} \mbox{kg}
\end{equation}
The value given by eq.~(\ref{eq:7a}) can be obtained by an independent
mode, based on the entropy consideration, too.

It has been rationalized by Bekenstein [7] that the entropy of a black hole 
can be expressed as 
\begin{equation}
\label{eq7}
S_{BH} \approx k.{\frac{{r_{BH}^{2}}} {{l_{Pc}^{2}}} } \approx 
k.10^{76}.\left( {{\frac{{m_{BH}}} {{m_{Sun}}} }} \right)^{2}
\end{equation}
where $k$ is the Boltzmann constant, $l_{Pc} $ is the Planck length, and 
$m_{Sun} $ is the current mass of our Sun. The second law of thermodynamics 
stipulates that for the irreversible processes (creation of a black hole may 
set as an example of such processes), the initial state (e.g., star) entropy 
must be lower than that of the final state (black hole). Suppose, a black 
hole creation is a consequence of gravitational collapse of a star with the 
mass $m_{Star} $. The entropy value of the magnitude of Boltzmann constant 
may be allocated to each degree of freedom of a classical system of 
particles (for the sake of simplicity, nucleons are considered). For the 
entropy of the initial system it must, therefore, hold
\begin{equation}
\label{eq8}
S \cong k.{\frac{{m_{Star}}} {{m_{n}}} } \cong k.10^{57}{\frac{{m_{Star} 
}}{{m_{Sun}}} }
\end{equation}
where $m_{n} $ is the mass of nucleon (usually the proton mass). 

Comparing (\ref{eq7}) and (\ref{eq8}) it can be seen that the process
of a collapse accompanied by a black hole creation is associated with
an increase in entropy by 19 orders of magnitude and it is, therefore,
of a strongly irreversible nature. It follows further from comparison
of (\ref{eq7}) and (\ref{eq8}) that reducing the initial mass
$m_{Star} $, the entropy of a corresponding black hole will drop by a
higher extent than that of initial star. Both the value become equal
at
\begin{equation}
\label{eq:10a}
m_{Star} \cong 10^{12} \mbox{kg}
\end{equation}
It means that at the time being no black holes with a lower mass can be 
created (and probably no such a black hole can exist). Otherwise the entropy 
of a black hole would be lower than that of the collapsing star, i.e. the 
second law of thermodynamics would be violated. 

It is worth mentioning that relations~(\ref{eq:7a}) and~(\ref{eq:10a})
have been obtained in independent ways.

In addition to the lowest mass limit, using (\ref{eq6}) the
gravitational radius of such a black hole, $r_{BH(\min )} $ can be
calculated
\begin{equation}
\label{eq9}
r_{BH(\min )} = \left( {l_{Pc}^{2} .a} \right)^{1 / 3}
\end{equation}
Introducing the present value of $a$ it is obtained
\begin{equation}
\label{eq:12a}
r_{BH(\min )} \cong 10^{ - 15} \mbox{m}
\end{equation}
which represents a nucleus dimension.

\section{The lowest mass of particles exerting gravitational influence}

Gravitational force is a far-reaching force with ostensibly unlimited
range.  Due to the existence of hierarchic rotational gravitational
systems, the range is, however, actually finite. This is a reason for
introducing so called ``effective gravitational range" $r_{ef} $, i.e.
the distance at which the density of gravitational field of a given
body is equal to the critical density of background gravitational
field. Critical energy density is in ENU equal to the actual mean
energy density, it decreases in time and is given [2] by
\begin{equation}
\label{eq10}
\varepsilon _{crit} = {\frac{{3c^{4}}}{{8\pi .G.a^{2}}}}
\end{equation}
It follows from (\ref{eq2}) and (\ref{eq10}) that (in absolute values)
\begin{equation}
\label{eq11}
{\frac{{3c^{4}}}{{8\pi .G.a^{2}}}} = {\frac{{3m.c^{2}}}{{4\pi .a.r^{2}}}}
\end{equation}
and, in turn
\begin{equation}
\label{eq12}
r_{ef} = \left( {r_{g} .a} \right)^{1 / 2}
\end{equation}
where $r_{ef} $ is the effective gravitational range of a body with the 
gravitational radius $r_{g} $. 

Based on the above rationalization, it is possible to determine the lightest 
particle able to exerts gravitational influence on its surroundings. The 
particle has the mass $m_{x} $ and its gravitational range is identical to 
its Compton's wavelength. Stemming from the following relation
\begin{equation}
\label{eq13}
\left( {{\frac{{2G.m_{x} .a}}{{c^{2}}}}} \right)^{1 / 2} = {\frac{{\hbar 
}}{{m_{x} .c}}}
\end{equation}
the current mass of the particle is
\begin{equation}
\label{eq:17a}
m_{x} \cong \left( {{\frac{{\hbar ^{2}}}{{2G.a}}}} \right)^{1 / 3} \cong 
10^{ - 28} \mbox{kg}
\end{equation}
and its Compton's wavelength
\begin{equation}
\label{eq:18a}
\lambda _{x} = {\frac{{\hbar}} {{m_{x} .c}}} = \left( {l_{Pc}^{2} .a} 
\right)^{1 / 3} \cong 10^{ - 15} \mbox{m}
\end{equation}
As follows from~(\ref{eq:12a}), it is at the same time also a lower
mass limit of a black hole. Turning back to entropy considerations, it
is obvious that relation (\ref{eq8}) should be modified as follows
\begin{equation}
\label{eq14}
S \cong k.{\frac{{m_{Star}}} {{m_{x}}} }
\end{equation}
In such a way, relations (\ref{eq6}), (\ref{eq:7a}), (\ref{eq8}),
and~(\ref{eq:10a}) become consistent since the entropy in (\ref{eq8})
will be time-dependent. Stemming from (\ref{eq12}), an increase of the
gravitational range of nucleons in time is an expected phenomenon.
Similarly, the degree of freedom and entropy of the system will
increase too. It should be pointed out that at the beginning of the
matter era it had to hold
\begin{equation}
\label{eq15}
m_{x} \cong m_{n} 
\end{equation}
It means that nucleons started to exert their gravitational influence
to the environment just at the beginning of the matter era which,
inter alia, allowed black holes to be formed. Stemming from the
equality of (\ref{eq7}) and (\ref{eq14}), it is possible to obtain
relations (\ref{eq6}) and~(\ref{eq:10a}).

\section{Upper mass limit of black holes}

Contrary to all the time present gravitational quanta, there were no 
elementary particles present the at the beginning of the Universe expansion. 
This is why it is useful to express the total entropy of the Universe, 
$S_{U} $ by a number of gravitational quanta [8] as follows
\begin{equation}
\label{eq16}
S_{U} = {\frac{{m_{U} .c^{2}}}{{{\left| {E_{g}}  \right|}}}} = 
{\frac{{c^{5}.t_{c}}} {{2G{\left| {E_{g}}  \right|}}}}
\end{equation}
where $m_{U} $ is the Universe mass and $E_{g} $ is the mean energy of the 
gravitational field quantum. The wavefunction of the mean gravitational 
quantum of the Universe is given by [8]
\begin{equation}
\label{eq17}
\Psi _{g} = \exp {\left[ {i.t\left( {t_{Pc} .t_{c}}  \right)^{ - 1 / 2}} 
\right]}
\end{equation}
where $t_{Pc} $ is the Planck time. Relation (\ref{eq17}) complies
with a Schr\"{o}dinger-like equation for the energy of gravitational
quanta $E_{g} $
\begin{equation}
\label{eq18}
E_{g} .\Psi _{g} = i.\hbar {\frac{{d\Psi _{g}}} {{dt}}}
\end{equation}
and it comes from (\ref{eq17}) and (\ref{eq18}) that
\begin{equation}
\label{eq19}
{\left| {E_{g}}  \right|} = {\frac{{\hbar}} {{\left( {t_{Pc} .t_{c}}  
\right)^{1 / 2}}}}
\end{equation}
Based on (\ref{eq16}) and (\ref{eq19}) for the total entropy of the
Universe it follows
\begin{equation}
\label{eq20}
S_{U} = \left( {{\frac{{a}}{{l_{Pc}}} }} \right)^{3 / 2} \cong 10^{92}
\end{equation}
No body, black holes including, can possess an entropy content higher
than that of the whole Universe. This is the reason due to which using
(\ref{eq7}) and (\ref{eq20}) it must hold
\begin{equation}
\label{eq21}
{\frac{{r_{BH(\max )}^{2}}} {{l_{Pc}^{2}}} } = \left( {{\frac{{a}}{{l_{Pc} 
}}}} \right)^{3 / 2}
\end{equation}
In (\ref{eq21}), $r_{BH(\max )} $ is the maximum gravitational radius
of a black hole in the Universe with the gauge factor $a.$ It can be
directly obtained from (\ref{eq21}) that
\begin{equation}
\label{eq22}
r_{BH(\max )} = \left( {a^{3}.l_{Pc}}  \right)^{1 / 4}
\end{equation}
which currently approaches to
\begin{equation}
\label{eq:28a}
r_{BH(\max )} \cong 10^{11} \mbox{m}
\end{equation}
The maximum mass of black holes, corresponding to~(\ref{eq:28a}) is then
\begin{equation}
\label{eq23}
m_{BH(\max )} \cong 10^{38}kg \cong 10^{8}m_{Sun} 
\end{equation}
which is in excellent agreement with experimental observation [9].

\subsection*{Acknowledgements}

The financial support of the research by the Slovak Grant Agency (Projects 
VEGA/1/6106/99) is appreciated.

\section*{References}

\begin{description}
\item {}[1] S.W. Hawking, Sci. Amer., 236 (1977) 34

\item {}[2] V. Skalsk\'{y}, M. S\'{u}ken\'{\i}k, Astrophys. Space Sci., 178 (1991) 
169

\item {}[3] V. Skalsk\'{y}, M. S\'{u}ken\'{\i}k, Astrophys. Space Sci., 181 (1991) 
153

\item {}[4] J. \v{S}ima, M. S\'{u}ken\'{\i}k, gr-qc/9903090

\item {}[5] P.C. Vaidya, Proc. Indian Acad. Sci., A33 (1951) 264

\item {}[6] M. S\'{u}ken\'{\i}k, J. \v{S}ima, gr-qc/0101026; Acta Phys. Pol., 
submitted

\item {}[7] J. Bekenstein, Phys. Today, 33 (1980) 24

\item {}[8] J. \v{S}ima, M. S\'{u}ken\'{\i}k, gr-qc/0103028; Entropy, submitted

\item {}[9] D.A. Kirzhnits, Czech. J. Phys., 48 (1998) 19
\end{description}

\end{document}